\documentclass[useAMS,usenatbib]{mn2e}
\usepackage{color}
\usepackage{amssymb}
\usepackage{amsmath}
\usepackage[dvips]{graphicx}
\usepackage{bm}
\usepackage{multirow}

\title{ Episodic Model For Star Formation History and Chemical Abundances in Giant and Dwarf Galaxies}

\author[Suma Debsarma, Tanuka Chattopadhyay, Sukanta Das, Daniel Pfenniger]
       {Suma Debsarma$^1$, Tanuka Chattopadhyay$^1$, Sukanta Das$^1$, Daniel Pfenniger$^2$ \\
         $^1$ Department of Applied Mathematics, University of Calcutta, 92
         A.P.C Road, Kolkata 700009, India; tanuka@associates.iucaa.in, \\
         $^2$ Observatoire de Gen\`eve, University of Geneva, CH 1290, Sauverny,  Switzerland;
          daniel.pfenniger@unige.ch }

\date{}
\pagerange{\pageref{firstpage}--\pageref{lastpage}} \pubyear{2012}

\begin{document}

\label{firstpage}

\maketitle

\begin{abstract}
  In search for a synthetic understanding, a scenario for the evolution of the star formation rate and the chemical abundances in
  galaxies is proposed, combining gas infall from galactic halos, outflow of gas by supernova explosions, and an oscillatory star
  formation process. The oscillatory star formation model is a consequence of the modelling of the fractional masses changes of the
  hot, warm and cold components of the interstellar medium. The observed periods of oscillation vary in the range
  $(0.1-3.0)\times10^{7}$\,yr depending on various parameters existing from giant to dwarf galaxies. The evolution of metallicity
  varies in giant and dwarf galaxies and depends on the outflow process. Observed abundances in dwarf galaxies can be reproduced
  under fast outflow together with slow evaporation of cold gases into hot gas whereas slow outflow and fast evaporation is
  preferred for giant galaxies. The variation of metallicities in dwarf galaxies supports the fact that low rate of SNII production
  in dwarf galaxies is responsible for variation in metallicity in dwarf galaxies of similar masses as suggested by various authors.
\end{abstract}

\begin{keywords}
Galaxies: star formation, Stars: winds outflows

\end{keywords}

\section{Introduction}
The delineation of star formation history (SFH) in galaxies has
become an increasingly powerful method for shaping galaxy
formation and evolution. For our own Galaxy we can observe that a
large number of stars and SFH is directly inferred from the age
distribution of stars. For distant galaxies the Hubble Space
Telescope is a powerful tool to resolve stellar populations into
SFH (Gallert et al. 2005). Star forming galaxies that sample the
faint end of galaxy luminosity function are not well studied like
brighter, more massive galaxies due to completeness limits of
large galaxy surveys (e.g., SDSS). Several authors (Lee et al.
2007; Kennicutt et al. 2008) have found marked differences in
H$_{\alpha}$ equivalent width (EW) measurements, normalized star
formation rate (SFR) over the past 5\,Myr, for star forming
galaxies fainter than $M_{B}\sim -15$ in comparison to brighter
galaxies. Different physical mechanisms like stochastic effects
(Mueller \& Arnett 1976; Gerola \& Seiden 1978), internal feedback
(stellar winds, supernova; e.g.\ Pelupessy et al.\ 2004; Stinson
et al.\ 2007), galaxy interactions (merger, tidal influences;
e.g.\ Toomre \& Toomre 1972; Gnedin 1999) have been proposed to
shed light in each of such processes. Also there are observational
differences in birth rate parameters ($b$), fraction of stars
formed per time interval ($f$), spatial distribution of stellar
components in dwarf galaxies (Weisz et al.\ 2008). There are
differences in duty cycles (the time for which the SFR is higher
than a threshold value) for giants and dwarfs also (Jaacks et al.\
2012; Lee et al.\ 2009) in case of episodic star formation
history.

A number of papers have investigated the formation (Burkert et
al.\ 1992; Chattopadhyay et al.\ 2012; Chattopadhyay et al.\ 2009;
Chattopadhyay \& Chattopadhyay 2007) and chemical evolution
(Matteucci \& Francois 1989; Amarsi et al.\ 2014; Homma et al.\
2015; Zahid 2014; de Boer et al.\ 2012) of the galaxies and other
spiral galaxies (Lynden Bell 1975; Sommer-Larsen 1991). Formation
of galactic disc through gas infall from the halo has been
suggested also by many authors (Twarog 1980; Hirashita et al.\
2001; Narayanan et al.\ 2015; Aumer et al.\ 2010; Combes 2008;
Haywood et al.\ 2015). Kennicutt et al.\ (1994) have suggested
that episodic star formation history and the star formation
activities differ for early and late type galaxies. Episodic star
formation has been suggested by de Boer et al.\ (2012), Nichols et
al.\ (2012) and many others.  All the observational as well as
theoretical studies demand an appropriate modelling of the SFH and
chemical history of the giant as well as dwarf galaxies taking
into consideration most of the physical mechanisms like infall,
feedback etc prevailing in these galaxies.

In the present work we give a synthetic model for the episodic
star formation in giant as well as in dwarf galaxies through a
fast and slowly damping cyclic scenario. For modelling the
chemical history we consider infall from halos of galaxies
together with outflow of gas due to supernova explosions. We
assume that the amount of gas mass driven out by the supernovae in
the form of wind is proportional to the instantaneous star
formation rate neglecting the time delay of $\lesssim 10^{7}$
years between the star formation and subsequent supernova
explosions (Samui 2014). Section 2 describes the mathematical
model. In section 3 we discuss about the values of the parameters
chosen for our study. Results and discussions are given in
sections 4 and 5. Section 6 outlines the main conclusions.

\section{Mathematical Model}
\subsection{Dynamical model of star formation in interstellar
medium}

The present mathematical model is mainly based on the model
suggested by Ikeuchi \& Tomita (1983) (hereafter IT83) which was
subsequently improved by Hirashita \& Kamaya (2000) and Hirashita
et al.\ (2001). According to the above model, star formation in a
giant galaxy takes place in the gaseous component consisting of
three parts, the hot, warm, and cold gas with fractional masses
$X_\mathrm{h}$, $X_\mathrm{w}$, $X_\mathrm{c}$ respectively. The
relative abundances of the above mentioned fractional masses are
controlled by supernova remnants (SNR) through the following
processes, already included in IT83: (i) sweeping of warm gas into
cold component ($aX_\mathrm{w}$, $a \sim
5\times10^{-8}$\,yr$^{-1}$), (ii) evaporation of cold clouds
embedded in hot gas ($bX_\mathrm{c}X_\mathrm{h}^2$, $b \sim
10^{-7}-10^{-8}$\,yr$^{-1}$), (iii) cooling of hot gas in warm gas
($cX_\mathrm{w}X_\mathrm{h}$, $c \sim
10^{-6}-10^{-7}$\,yr$^{-1}$).  In the present work we have
included one more process, which is the sweeping of hot gas into
cold gas ($fX_\mathrm{h}$). Then the corresponding time rate of
change of the fractional masses reduce to,
\begin{eqnarray}
  \frac{dX_\mathrm{c}}{dt} &=&  a X_\mathrm{w} - bX_\mathrm{c} X_\mathrm{h}^2 + f X_\mathrm{h} ,\\
  \frac{dX_\mathrm{w}}{dt} &=& \!\!\!\!\!-a X_\mathrm{w} + c X_\mathrm{w}X_\mathrm{h} ,\\
  \frac{dX_\mathrm{h}}{dt} &=&  b X_\mathrm{c}X_\mathrm{h}^2 - c X_\mathrm{w}X_\mathrm{h} - f X_\mathrm{h},
\end{eqnarray}
where the coefficients  $a$, $b$, $c$, and $f$ are constants.

We normalize the time $t$ and the coefficients as, $\tau=ct$,
$A=a/c$, $B=b/c$, $F=f/c$ and normalize the mass fractions by
$X_\mathrm{c}+X_\mathrm{h}+X_\mathrm{w}=1$.  Then the system of
three differential equations reduces to two.  Substituting $x =
X_\mathrm{c}$, and $ y = X_\mathrm{h}$, we have
\begin{eqnarray}
\label{eq:dynsysa}
  \frac{dx}{d\tau} &=&  A (1-x-y) - B xy^2 + F y,\\
\label{eq:dynsysb}
  \frac{dy}{d\tau} &=& -y (1-x-y) + B xy^2 - F y.
\end{eqnarray}
The above set of equations can be treated as a system of coupled
polynomial ordinary differential equations including up to third
degree terms.  The fractional mass condition limits the phase
space to the triangular domain $x \geq 0$, $y\geq0$, $x+y\leq 1$.

We study the phase space structure first through its stationary
states, or fixed points, which are derived in detail in the
Appendix:
\begin{enumerate}
\item $\left\{x=1,y=0\right\}$, \item
$\left\{x=\frac{1-A+F}{1+AB},y=A\right\}$, \item $\left\{x={1
\over 2} - \sqrt{{1\over 4} - F/B}, y={1 \over 2} +
  \sqrt{{1\over 4} - F/B}\right\}$,
\item $\left\{x={1 \over 2} + \sqrt{{1\over 4} - F/B}, y={1 \over
2} -
  \sqrt{{1\over 4} - F/B}\right\}$.
\end{enumerate}
Solution (i) exists for any value of $A$, $B$, and $F$, and does
not depend on them.  This is an extreme case where all the mass is
in the cold phase $X_\mathrm{c}=1$.

Solution (ii) depends on all three parameters $A$, $B$, and $F$,
and admits values of $X_\mathrm{c}$, $X_\mathrm{w}$, and
$X_\mathrm{h}$ in the interval $[0-1]$ for suitable positive $A$,
$B$, and $F$, subject to restrictions, such as $F\leq AB(1-A)$,
and $0\leq A \leq 1$.

Solutions (iii), and (iv) depend only on $B$, and $F$.  Since for
each of them $x+y=1$, the warm phase fraction vanishes,
$X_\mathrm{w} = 0$, for any value of $B$, and $F$.  To exist these
solutions need to be real, which gives a constraint on the
parameters, $B\geq 4F$.

In case of giant molecular clouds $ X_\mathrm{c}$, $X_\mathrm{h}$,
$X_\mathrm{w} $ should be all positive to some extent, the cases
where $X_\mathrm{w}$ vanishes, or both $X_\mathrm{w}$ and
$X_\mathrm{h}$ vanish may be thought as some kind of exceptional
limiting conditions where proper mixing or supernova explosions
have not taken place. Therefore solution (ii) is the only
stationary point representing a realistic stationary situation for
star forming clouds.

Also it follows (Appendix) that for $ 0 < A < 1.0$, $0< F< AB(1 - A)$
point (ii) moves within the triangle $ x \ge 0, y \ge 0$ and $x + y <1
$ (Figs.~1--4). The figures indicate that the velocity field may or
may not rotate about the stationary point (ii) only.  For particular
parameters, such as $A=0.4$, $B=2.5$, $F=0.25$ there exists a limit
cycle (Fig.\ 4), which means that the system converges toward a
periodic attractor displaying oscillations of its three component
abundances and its star formation rate.

Examining the eigenvalues as a function of the parameters ($0 < A <
1.0$, $0 < B < 100$, $0 \leq F < AB(1-A)$) show that for (ii) (see
Fig.~5) the real part is mostly negative and in case of lower values
of $B$ the real part is less negative. This means that when $B$ is
small, evaporation rate of cold gas to hot gas is less. Lower rate of
evaporation means less abundance of hot gas, i.e., lower rate of
supernova production. This happens in case of dwarf galaxies. Thus the
episodic star formation process prevails for a longer time in dwarf
galaxies compared to giant galaxies. For particular parameters,
such as $A=0.4$, $B=2.5$ and $F=0.25$ there exists a limit cycle. This
means dwarf galaxies are the favourable places for episodic star
formation. On the other hand higher values of $B$ is more likely for
giant galaxies. Hence episodic star formation process decays rapidly
in giant galaxies (viz. Figs.~6, 7). Perhaps this is the reason why
giant galaxies do not show evidences of episodic star formation in the
last 5\,Myr compared to dwarfs (Lee et al.\ 2007; Kennicutt et
al.\ 2008).

Point (i) is a stable focus (Fig.~8), point (iii) is stable or
unstable hyperbolic (Fig.~9) and point (iv) is unstable
hyperbolic or stable spiral (Fig.~10). None of these cases
provide conditions for episodic star formation.

Thus Figs.~5, 8--10 give a clear demonstration of the existence of
cyclic star formation as a function of the various parameters $A$,
$B$, $F$. These parameters are representatives of the physical
processes occurring in galaxies. Lower values of these parameters
correspond to the situation favourable for dwarfs and vice versa
as discussed above. In Fig.~5, the real part of $\lambda$ is less
negative or zero and the non-zero imaginary part represents an
elliptic orbit, which means there exists cycles of slowly
decreasing or constant amplitudes for lower values of $A$, $B$,
and $F$ whereas rapidly decreasing amplitudes for higher values of
$A$, $B$, and $F$. This indicates that around the stationary point
(ii) the gas abundances participating in star formation rotates in
abundance space either with a slowly decaying manner or continue
rotation in case of a limit cycle in dwarfs but always with a
rapidly decaying manner in giants. On the other hand for
stationary point (i) Fig.~8 shows that $\lambda$ is always
negative with no imaginary part, i.e., there exists no cycles for
$A$, $B$, and $F$. Figs.~9, 10 indicate that the real part of
$\lambda$ has both positive as well as negative values, i.e., the
mode of vibration might be stable or unstable. Hence for
stationary points other than (ii) there are no cyclic variations
in abundance space for any values of the parameters $A$, $B$, and
$F$.

\subsection{Chemical evolution in galaxies}
The chemical evolution in galaxies is based on two phenomena: (i)
The gradual gas infall from halos, and (ii) gas outflow as a
result of supernova explosion. We have not considered the above
two phenomena in the dynamical model of star formation because the
infall time scale ($t_\mathrm{in}$) is much larger than the
oscillatory star formation time scale (viz.\ average duty cycles,
Tables 4, 5). In clusters of galaxies, galaxies move through the
intracluster medium and if the ram pressure force exceeds the
internal gravitational pull, gas will be stripped from these
galaxies (Irwin et al.\ 1987; White et al.\ 1991; B\"{o}hringer et
al.\ 1995; Balsara et al.\ 1994). Ram pressure stripping is more
efficient in dwarf galaxies compared to giant galaxies due to its
low potential well. The time scale of gas removal from dwarf
galaxies ($M\lesssim 10^9 M_{\bigodot}$) is of the order of
$2.02\times10^8$\,yr and $2.19 \times10^9$\,yr for massive dwarf
galaxies (Mori \& Burkert 2001). This is also larger than the
oscillatory star formation time scale ($\sim 10^7$\,yr viz.\ duty
cycles, Tables 4, 5). So the oscillatory model of star formation
discussed in section 2.1 does not include the effect of infall as
well as outflow.

The changing rates of gas mass ($M_\mathrm{g}$) and metal mass
($M_i$, $i= $Fe, O, etc.) are mainly based on the work done by
Hirashita et al.\ (2001) but in our case we have included outflow
due to supernova explosions. Massive stars would explode as a
supernova after $10^7$\,yr and these explosions drive cold gas out
of the galaxy as galactic wind (viz.\ ram pressure stripping). We
assume that the gas mass driven is proportional to the
instantaneous star formation rate neglecting the time delay
between star formation and subsequent explosion. Then the
evolution of gas mass ($M_\mathrm{g}$) and metal mass
($M_\mathrm{i}$) are governed by the set of differential
equations,
\begin{eqnarray}\label{eq:Mg}
  \frac{dM_\mathrm{g}}{dt}&=&-\psi(t)+R_\mathrm{ins}\psi(t)+I(t)-\eta_\mathrm{w}\psi(t),
\end{eqnarray}
\begin{equation}
  \frac{dM_\mathrm{i}}{dt}=-X_i(t)\psi(t)+\left(R_\mathrm{ins}X_i(t)+Y_{i,\mathrm{ins}}\right)\psi(t)
     + I(t)X_i^\mathrm{f}
\end{equation}
where $R_\mathrm{ins}$ and $Y_{i,\mathrm{ins}}$ are the instantaneous
returned fraction of gas from stars and the fractional mass of the
newly formed element $i$ respectively.  $X_i$ is the abundance of
$i$'th element ($X_i\equiv M_i/M_\mathrm{g}$) and $X_i^\mathrm{f}$ is
the abundance of $i$'th element in the infall material (Tinsley 1980).
$\eta_\mathrm{w}\psi(t)$ is the outflow and $\eta_\mathrm{w}$ is a
proportionality constant defined as (Samui 2014)
\begin{equation}
\eta_\mathrm{w}=\left(\frac{v_\mathrm{c}}{v_\mathrm{c}^0}\right)^{-\alpha}
\end{equation}
where $v_\mathrm{c}$ is the rotational velocity of the galaxy and
$v_\mathrm{c}^0$ is that when $\eta_\mathrm{w}=1$. $\alpha=2$ or 1
on whether the outflows are energy driven or momentum driven. It
is clear that for a given $\psi(t)$, outflow is higher for a dwarf
galaxy than for a giant galaxy. $I(t)$ is the rate of gas infall
from halo at time $t$ and is given by (Hirashita et al.\ 2001)
\begin{equation}
I(t)=\frac{M_0}{t_\mathrm{in}} \exp(-t/t_\mathrm{in})
\end{equation}
where $M_0$ is the total mass of infall, $t_\mathrm{in}$ is the
total time of infall. $\psi(t)$ is the star formation rate at time
$t$ and is given by (Hirashita et al.\ 2001),
\begin{equation}
\psi(t)=M_\mathrm{g}x/t_*
\end{equation}
Dividing Eqs. (\ref{eq:Mg}) and (7) by $M_0$ and substituting
$f_\mathrm{g}=\frac{M_\mathrm{g}}{M_0}$,
$\widetilde{\psi}=\frac{\psi}{M_0}$ as done by Hirashita et al.\
(2001) we have
\begin{eqnarray}
\frac{df_\mathrm{g}}{dt} &=&
-(1-R_\mathrm{ins})\widetilde{\psi}(t) +
\frac{1}{t_\mathrm{in}} \exp(-t/t_\mathrm{in}) \nonumber \\
&& \qquad\qquad\qquad\qquad
  -\left(\frac{v_\mathrm{c}}{v_\mathrm{c}^0}\right)^{-\alpha}\widetilde{\psi}(t),\\
  \ f_\mathrm{g}\frac{dX_i}{dt} &=& Y_{i,\mathrm{ins}}\widetilde{\psi}(t)
  - \frac{X_i(t)-X_i^\mathrm{f}}{t_\mathrm{in}} \exp(-t/t_\mathrm{in}) \nonumber \\
&& \qquad\qquad\qquad\qquad
  + X_i(t)\left(\frac{v_\mathrm{c}}{v_\mathrm{c}^0}\right)^{-\alpha}\widetilde{\psi}(t)
\end{eqnarray}

\section{Initial values of the parameters}
We have chosen the values of the parameters $Y_{i,\mathrm{ins}}$,
$X_i^\mathrm{f}$, $R_\mathrm{ins}$, $t_*$, $t_\mathrm{in}$ from
Hirashita et al.\ (2001).  We have chosen the values of $A$, $B$
and $F$ as discussed in section 2.1. Samui (2014) has discussed on
$v_\mathrm{c}, v_\mathrm{c}^0$ and $\alpha$. For dwarf galaxies
$v_\mathrm{c}$ varies from $10\rm\,km s^{-1}$ to $30 \rm\,km
s^{-1}$ whereas for giants the range is from $100 \rm\,km s^{-1}$
to $800 \rm\,km s^{-1}$. The parameter $\alpha$ varies from 1 to
2. Hence we have chosen the parameters accordingly so that they
can produce observed metallicities for giant as well as dwarf
galaxies. All the values of the parameters considered are given in
Table 2.

\section{Results}
In the present work we have tried to model an episodic star
formation history and metal production observed in giant as well
as in dwarf galaxies. For the dynamical model we have included an
additional mechanism of sweeping of hot gas into cold gas besides
sweeping of warm gas into cold gas to represent a more realistic
situation of giant molecular clouds. Also in the model of metal
evolution we have added outflow of gas from galaxies due to
supernova explosions. This helps to study the evolution of
metallicity in giant as well as in dwarf galaxies under various
parametric conditions.  While studying the star formation rate
(SFR) we have found a stationary point (ii) around which a slow
elliptic inspiraling behaviour of the eigenvalue exists for lower
values of $B$ (Figs.~1, 2, 3, 5) whereas the process is fast for
higher values of $B$ (e.g., $B=25$ in Fig.~3). Under special
parameter condition there exists a limit cycle (Fig.~4 and
Appendix), e.g., $A = 0.4$, $B = 2.5$, $F = 0.25$. Tables 1, 3 show
the average star formation rates and abundances in giants and
dwarf galaxies for various values of $A$, $B$ and $F$. It is clear
that both $B$ and $F$ are lower and higher in dwarfs and giants
respectively to produce observed abundances whereas values of $A$
are similar for both kind. The average star formation rate is
higher in giants than in dwarfs. The above trend suggests that in
dwarfs due to lower rate of supernova production, mixing of hot
gas and evaporation of cold gas is lower but due to less amount of
gas the SFR is lower compared to giants where in spite of larger
evaporation ($B$ is higher) and fast mixing ($F$ is larger) the
SFR is higher due to higher potential well. The variation of metal
abundances in dwarfs of similar masses supports the results
derived by Kroupa (2004) that low rate of SNII production in dwarf
galaxies is responsible for the variation of metallicities in
dwarfs of similar masses.  This leads to the fact that galaxian
stellar initial mass function is very much influenced by the
galaxy masses which in turn affects the star formation history.
Cosmologically this implies that the number of SNII per low-mass
star is significantly depressed and that chemical enrichment
process is much slower in dwarf galaxies with a low average
star-formation rate compared to giant galaxies.

In Fig.~11 we have compared the predicted iron abundances at
different times for giant and dwarf galaxies with the observed
ones from Rocha-Pinto et al.\ (2000) and Chattopadhyay et al.\
(2012). The age-metallicity evolution gives good fit both for
giant as well as dwarf galaxies ($p$-values are 0.713 and 0.653
excluding few outliers, for giants and dwarfs respectively).
Tables 4 and 5 list the ratio of average star formation rates in
the last 100\,Myr, 500\,Myr, 1\,Gyr normalized to the life time
average star formation rate (viz.\ $b_{100}, b_{500}, b_{1000}$)
and average duty cycles for giant and dwarf galaxies for various
initial values of the parameters (viz.\ $A$, $B$, $F$) and they
are compared with the corresponding observed quantities for dwarf
irregular galaxies of M81 group (Weisz et al.\ 2008). The values
for dwarf galaxies are very close to the observed ones in most
cases. The differences might be due to the fact that Weisz et al.
(2008) considered only a particular type of dwarf galaxies e.g.
dwarf irregulars (dIrr) other than considering all species e.g.
dwarf ellipticals (dE) or dwarf spheroidals (dSph) under different
abundance properties. The average duty cycles for giant and dwarf
galaxies are in the ranges $(0.1 - 3.2)\cdot10^7$\,yr and
$(0.1-0.6)\cdot 10^7$\,yr respectively. The duty cycles for giants
are initially larger and rapidly die out whereas those for dwarfs
remain more or less similar.

\section{Discussion}
Eigenvalues plots as a function of the parameters shows that out
of the four stationary points (ii) is the only possible stable
stationary state for a three-phase medium which is stable for
particular values of the parameters.  For (ii) the real part of
complex eigenvalues when negative indicates that the
stationary point is stable, and the non-zero imaginary part indicates that
nearby solutions rotate around the point. When the real part of eigenvalue
is less negative, i.e., $B$ is small, the decay is slow and when
real part of eigenvalue is more negative, i.e., $B$ is larger, the
orbit decays faster. There exists a stable attractor only under
very special parametric situations, e.g.,  $A=0.4$, $B=2.5$,
$F=0.25$ where star formation remains episodic over the entire
period. We have already mentioned that smaller values of $B$ are
likely in dwarf galaxies and vice versa. Also limit cycle
behaviour exists for small $B$. Hence episodic star formation has
frequent occurrence in dwarf galaxies for all the time and for the
initial phases only in giant galaxies. Thus in giant galaxies
other processes of star formation is preferred (e.g., galaxy
interaction etc.) over episodic process.

The model and results indicate that the observed metal abundances
for giants and dwarfs are reproduced for larger and smaller values
of $B$ respectively. From Tables 1, 3 we find that in dwarf
galaxies, due to low potential well, the amount of gas available
for star formation is much less compared to giant galaxies. Hence
rate of supernova production is less in dwarfs compared to giants.
As a result evaporation rate of cold gas into hot gas is low.
Since SFR is proportional to the amount of cold gas, average SFR
is lower in (viz.\ Table 1, column 4) these dwarfs due to low
potential well than in giant galaxies (viz.\ Table 3, column 4),
where $B$ is higher. Hence enough cold gas is available for star
formation in spite of higher evaporation rate due to higher
potential well resulting in a higher average star formation rates.
Since rate of supernova explosions is lower in dwarfs compared to
that of giants, heavy metals are less abundant in dwarfs rather
than giants (viz.\ Tables 1, 3, Figs.~11, 12). In our model
outflow has been considered, to be originated due to driven out of
hot gas as a result of supernova explosions. In such outflow it is
inversely proportional to the square of the circular velocity of
the galaxy (Samui et al.\ 2010). Now the circular velocity of
dwarf galaxies vary from $10\rm\,km\,s^{-1}$ to $30
\rm\,km\,s^{-1}$, which is very low compared to that of giant
galaxies.  Hence in dwarf galaxies the rate of outflow is much
higher than that of giant galaxies. This also reduces the process
of star formation and as a result the metal production is lower in
dwarfs than in giants. The birth rate of stars in the last
100\,yr, 500\,yr, 1000\,yr are more or less the same in dwarfs but
vary for giants. Hence it is expected that the starbursts are
more likely to happen in the initial phase of star formation
history in giant galaxies whereas the birth rates remain the same for
dwarfs over the entire star formation history.

\section{Conclusion}

In the present work an episodic star formation scenario has been
proposed for the observed star formation history in giant as well
as dwarf galaxies. The model is based on the transition of hot and
warm gases into cold gas and vice versa. The rate of transition
processes have been treated as parameters to study the behaviours
of observable quantities, e.g., duty cycles, birth rates of star
formations in the last couple of years, present metallicities in
giant as well as dwarf galaxies, and compared with observations.
The present model is an extension of the work by Hirashita et al.\
(2001) taking into additional consideration of (i) sweeping of hot
gas into cold gas and (ii) outflow of gas due to supernovae
explosion which in turn give a picture of episodic star formation
scenario in giant as well as dwarf galaxies. It is found that life
time average SFR of dwarfs are lower compared to giant galaxies
and the birth rate of stars in the last 100\,yr, 500\,yr, 1000\,yr
are more or less the same in dwarfs but vary for giants. Hence it is
expected that starbursts are more likely to happen in the initial
phase of star formation history in giant galaxies whereas the
birth rates remain same for dwarfs over the entire star formation
history. The average star formation cycles of dwarfs are smaller
than giants by a factor of almost 2 (viz.\ last columns of Tables
4 and 5). The observed metallicities are produced for a slow
sweeping, slow evaporation, and fast outflow of warm and hot gases
in dwarfs compared to fast sweeping, fast evaporation, and slow
outflow in giants which is compatible with the physical structure
of low and high galaxy potential of these respective galaxy types.
The variance of metallicities in dwarfs might be due to low rate
of SNII production as suggested by many authors.

\begin{appendix}
\section{Calculation and stability of the stationary points}

\subsection*{Stationary points:}
Setting the rhs of Eqs.\ (\ref{eq:dynsysa}) and (\ref{eq:dynsysb})
to zero we get,
\begin{eqnarray}
  \label{eq:zeroa}
  A(1-x-y)-Bxy^2+Fy &=& 0 , \\
  \label{eq:zerob}
  -y(1-x-y)+Bxy^2-Fy &=& 0.
\end{eqnarray}
Factoring (\ref{eq:zerob}), we obtain two possible constraints for
$y$,
\begin{equation}
  \label{eq:solyzeroa}
  y=0,
\end{equation}
or
\begin{equation}
  \label{eq:solyzerob}
  y=\frac{1+F-x}{1+Bx}.
\end{equation}
Substituting the first possibility $y=0$ in Eq.~(\ref{eq:zeroa})
we obtain $x=1$. Hence
\begin{equation}
  \label{eq:solxyone}
  (x=1,y=0)
\end{equation}
is a possible stationary point (viz. (i)). This solution exists
for any values of $A$, $B$, and $F$.

Substituting the second possibility Eq.~(\ref{eq:solyzerob}) in
Eq.~(\ref{eq:zeroa}), we obtain a cubic equation for $x$, which
factors as a product of a linear and quadratic terms,
\begin{equation}
  \label{eq:polynx}
   \left[(1+A B) x + A -F - 1\right]\left[B x(x-1) +F\right] = 0.
\end{equation}
Assuming the first term in brackets as zero, we obtain the stationary
point (ii),
\begin{equation}
  \label{eq:solxytwo}
  \left(x = {1-A+F \over 1+AB}, y = A\right),
\end{equation}
which depends on $A$, $B$, and $F$.  Since physical solutions
require $0\leq x,y\leq 1$, $0\leq x+y \leq 1$, $A\geq 0$, $B\geq
0$, and $F\geq 0$, for this stationary point to exist the
parameters must additionaly satisfy $0\leq A\leq 1$, and $F \leq
AB(1-A)$.

Assuming the second term in brackets in Eq.~(\ref{eq:polynx}) as zero, and
$B>0$, yields two more stationary points, (iii) and (iv),
\begin{equation} \label{eq:solxythreefour}
  \left( x = {1 \over 2}\left(1 \mp \sqrt{1 - {4F\over B}}\right),
         y = {1 \over 2}\left(1 \pm \sqrt{1 - {4F\over B}}\right)\ \right),
\end{equation}
which are real and distinct if $0< 4F\leq B$.  These points do not
depend on $A$ and lie on the diagonal $x+y=1$, which implies that
the third warm component vanishes, $X_\mathrm{w}=0$.

\subsection*{Stability of the stationary points:}
The Jacobian matrix $J$ of the dynamical system Eqs.\
(\ref{eq:dynsysa}) and (\ref{eq:dynsysb}) is,
\begin{equation}
  J=\left(
    \begin{array}{cc}
      -By^2 -A & -2Bxy - A + F \\
      By^2 + 1 & 2Bxy + x + 2y - F - 1
    \end{array}
  \right) .
\end{equation}
Hence, $|J-\lambda I|=0$ gives the characteristic polynomial,
\begin{equation}
  \label{eq:charpol}
  \lambda^2-P\lambda-Q=0,
\end{equation}
where,
\begin{equation} \label{eq:P}
P = -By^2 +2(Bx+1)y - A-F+x-1, \\
\end{equation}
\begin{equation} \label{eq:Q}
Q=2By^3 - B(A+x+1)y^2 + (2ABx+A+F)y + A(x-F-1) ,
\end{equation}
For stability, the real part of the eigenvalues must be negative
or zero. The eigenvalue of imaginary part indicates that nearby
solutions rotate around the stationary point in phase space.

\subsubsection*{Point (i)}
For stationary point (i) the eigenvalues are the solutions for
$\lambda$ of Eq.~(\ref{eq:charpol}), found to be,
\begin{equation}
  \lambda_1=-A, \quad \lambda_2=-F.
\end{equation}
Since physically $A,F > 0$, this point is always attractive and
stable.

\subsubsection*{Point (ii)}
For stationary point (ii), the eigenvalues are more complicated to
characterize.  Substituting Eq.~(\ref{eq:solxytwo}) in Eqs.~(\ref{eq:P}),
(\ref{eq:Q}), $P$ and $Q$ for point (ii) read
\begin{equation}\label{eq:PQtwo}
  P=\frac{1+F-A^2B-2A}{1+1/AB}, \quad Q=A(A^2B-AB+F).
\end{equation}
The discriminant $\Delta=P^2+4Q$ of Eq.~(\ref{eq:charpol}) reads,
\begin{equation}
  \Delta = {A S \over(AB+1)^2} ,
\end{equation}
where $S$ is a quadratic polynomial in $F$,
\begin{equation} \label{eq:quadF}
  S = c_2 F^2+c_1F +c_0,
\end{equation}
where
\begin{eqnarray}
  c_2 &=&  AB^2,\\
  c_1 &=& -2A^3B^3+2AB^2+8AB+4,\\
  c_0 &=&  AB(A^4B^3+8A^3B^2-6A^2B^2 + \nonumber\\
      &&   \qquad 12A^2B-12AB+4A+B-4)
\end{eqnarray}
The sign of $S$ is also the sign of $\Delta$, which tells whether the
eigenvalues are complex or real.  Solving Eq.~(\ref{eq:quadF}) for $F$
knowing $A$ and $B$ gives the limits in the $(A,B,F)$ space for real
or complex eigenvalues.  The exact boundary in parameter space can be
specified by solving multivariate high order polynomials in $A$, $B$,
and $F$ for the real and imaginary parts of the eigenvalues.  These
expressions are too large to show here.

But when the eigenvalues are complex, the sign of $P$ is also the sign
of the real part of both eigenvalues, which is also the sign of
$1+F-A^2B-2A$. Thus $B< (1+F-2A)/A^2$, or alternatively $F >
A^2B+2A-1$ when (ii) is unstable and complex.  As seen above,
$F < AB(1-A)$ for $(x,y)$ to be real, thus $A^2B+2A-1 < F < AB -A^2B $, or
$2A(AB+1) < AB+1$. Since $A$ and $B$ are positive, the additional
constraint $A<1/2$ follows.

Fig.~5 shows the real and imaginary part of the eigenvalues for
(ii) for $0 < A <1.0$, $B =3$, $0< F < 1.5$. The real parts
are slightly positive when $A<0.5$, $F<0.6$, while the
imaginary part is non-zero, which is a necessary but not
sufficient condition for the existence of a limit cycle in the
neighbourhood of the stationary point.

\subsubsection*{Points (iii) and (iv)}
For stationary points (iii) and (iv), the eigenvalues are
\begin{eqnarray}
  \lambda_{1\pm} &=& -A + {1\over2} \left(1 \pm \sqrt{1-{4F\over B}} \right), \\
  \lambda_{2\pm} &=& 2F - {B\over2} \left(1 \pm \sqrt{1-{4F\over B}} \right).
\end{eqnarray}
Since $B\geq 4F$ for the stationary points to exist, these eigenvalues
are real, and can be negative or positive. Interestingly, although
the positions of the stationary points are indepedent on $A$, the
first eigenvalue does depend on $A$.

Figs.~8, 9, and 10 show that points (i), (iii), and (iv) are all
stable or unstable, where the nearby orbits are hyperbolic, which
means that nearby orbits do not rotate around the points in phase
space.  This is expected since these stationary points are on the
boundary of the triangular physical domain, so nearby real
solutions cannot rotate around the stationary point.

\subsection*{Numerical solutions}

Over 1600 phase space plots and movies have been numerically computed for
$0 < A \leq 1$, $0 < B \leq 100$, and $0\leq F < AB(1-A)$ for the cases
where the stationary point (ii) is inside the physical triangle $ 0
\leq x \leq 1$, $0 \leq y \leq 1$ and $ x + y \leq 1$.%
\footnote{The phase space plots and movies are available at
  {\tt https:/obswww.unige.ch/$\sim$pfennige/DCDP/} in repsectively the subfolders
  {\tt PhaseSpacePlots} and {\tt Movies}.}
Each plot contains the vector field (in gray) and solutions starting
from $x=0$, $y = 0.05,0.10,\ldots ,0.95$ (in blue), and other
solutions starting very close to the four stationary points (in red),
which are visible only when the stationary points are unstable.  Each
solution is integrated up to a time value of 500.  In each plot the
parameters $A$, $B$, and $F$ are indicated, as well as the positions
$(x,y)$ of the four stationary points (i)--(iv), and their respective
real or complex eigenvalues.

These particular solutions can be used to check the existence of a
stable limit cycle (a periodic oscillating solution) around point
(ii).  A necessary but not sufficient condition for the existence of
such a stable periodic solution is that the point (ii) eigenvalues are
complex with positive real part. Further, solutions starting close to
some other unstable stationary points must asymptotically wind up
around point (ii) keeping a finite distance to it (see Fig.~4), which
is granted when point (ii) eigenvalues are complex with positive real
part.%
\footnote{The phase space plots with detected limit cycles are duplicated
in subsfolder {\tt LimitCycles}.}

When point (ii) eigenvalues are complex but with negative real part
all the solutions in its neighbourhood spiral toward point (ii).

The movies are sequences of phase space plots over the parameters
missing in the file name, varying $F$ first, $B$ second and $A$ last
in ascending order of the parameter values.

The found limit cycles are all consistant with the necessary parameter
constraints $A<1/2$, $B< (1+F-2A)/A^2$, and $F>A^2B+2A-1$ discussed
above.  For $F = 0$ the values of $A$ and $B$ are consistent with that
of Hirashita et al. (2001) for the stationary point (ii).

\end{appendix}

\clearpage
\begin{figure}
\centering
\includegraphics[width=16cm]{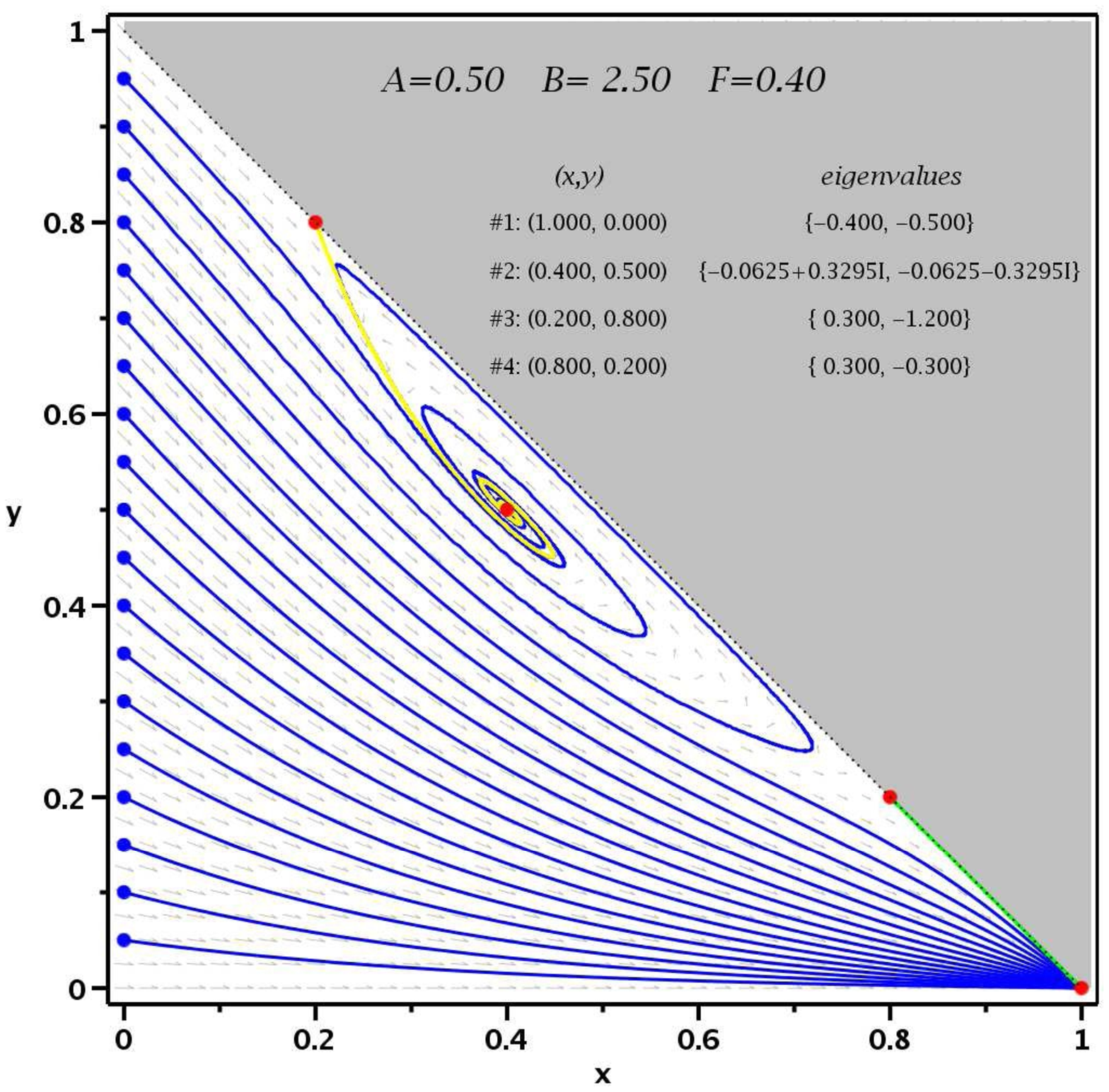}
\caption{Phase space plot for $A=0.5$, $B=2.5$, $F=0.4$.  The velocity
  field is indicated with grey arrows. The stationary points are
  marked by red dots, their respective $(x,y)$ coordinates are
  indicated, as well as their eigenvalues.  The blue curves are
  particular solutions starting at the positions indicated by blue
  dots. The yellow and green curves are particular solutions starting
  near unstable stationary point (iii) and (iv).  The complex and
  stable stationary point (ii) attracts nearby phase space with inward
  spiral motion.}
\label{Figure:1}
\end{figure}

\clearpage
\begin{figure}
\centering
\includegraphics[width=16cm]{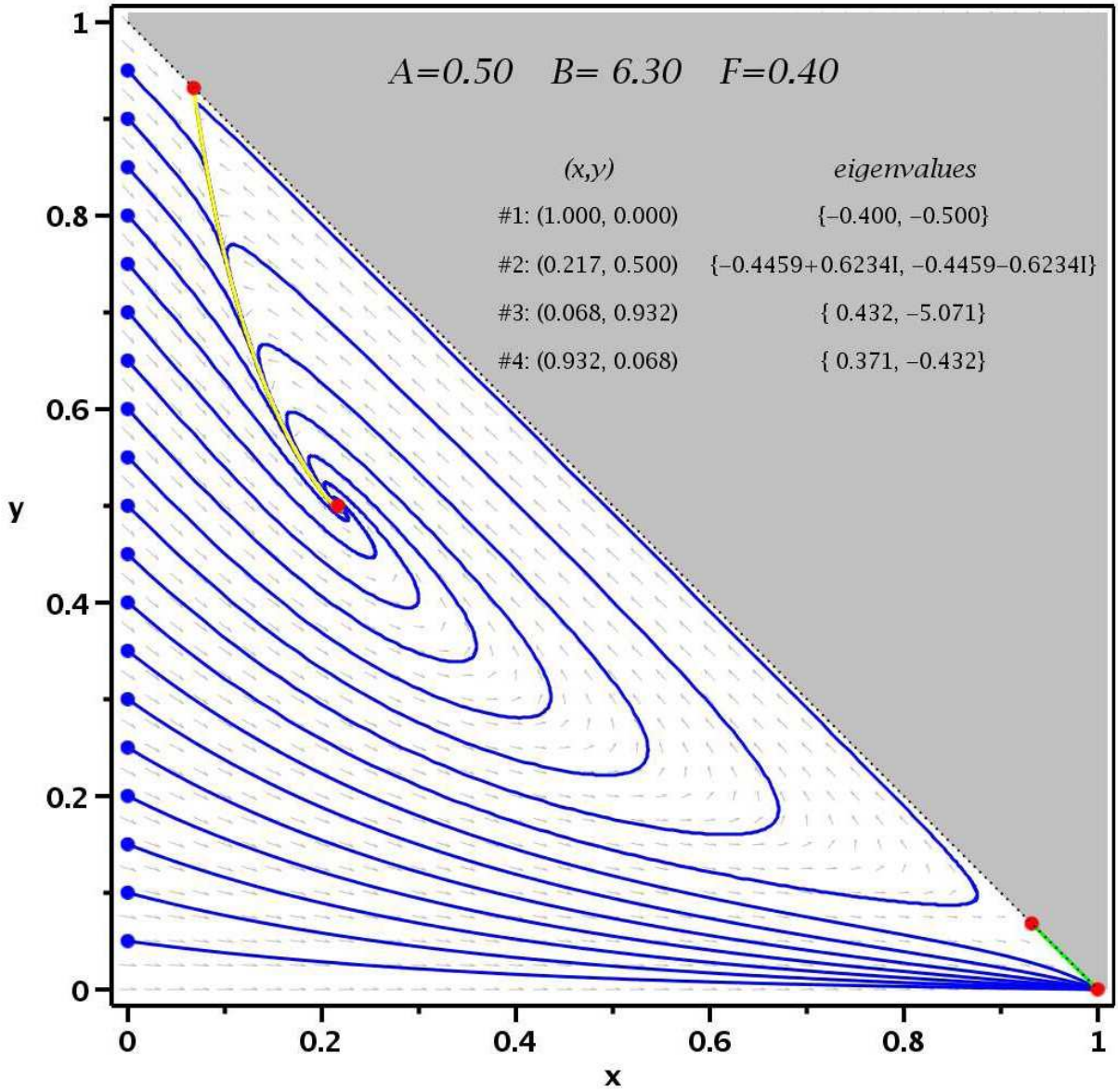}
\caption{Phase space plot for $A=0.5$, $B=6.3$, $F=0.4$, as in Fig.~1.}
\label{Figure:2}
\end{figure}

\clearpage
\begin{figure}
\centering
\includegraphics[width=16cm]{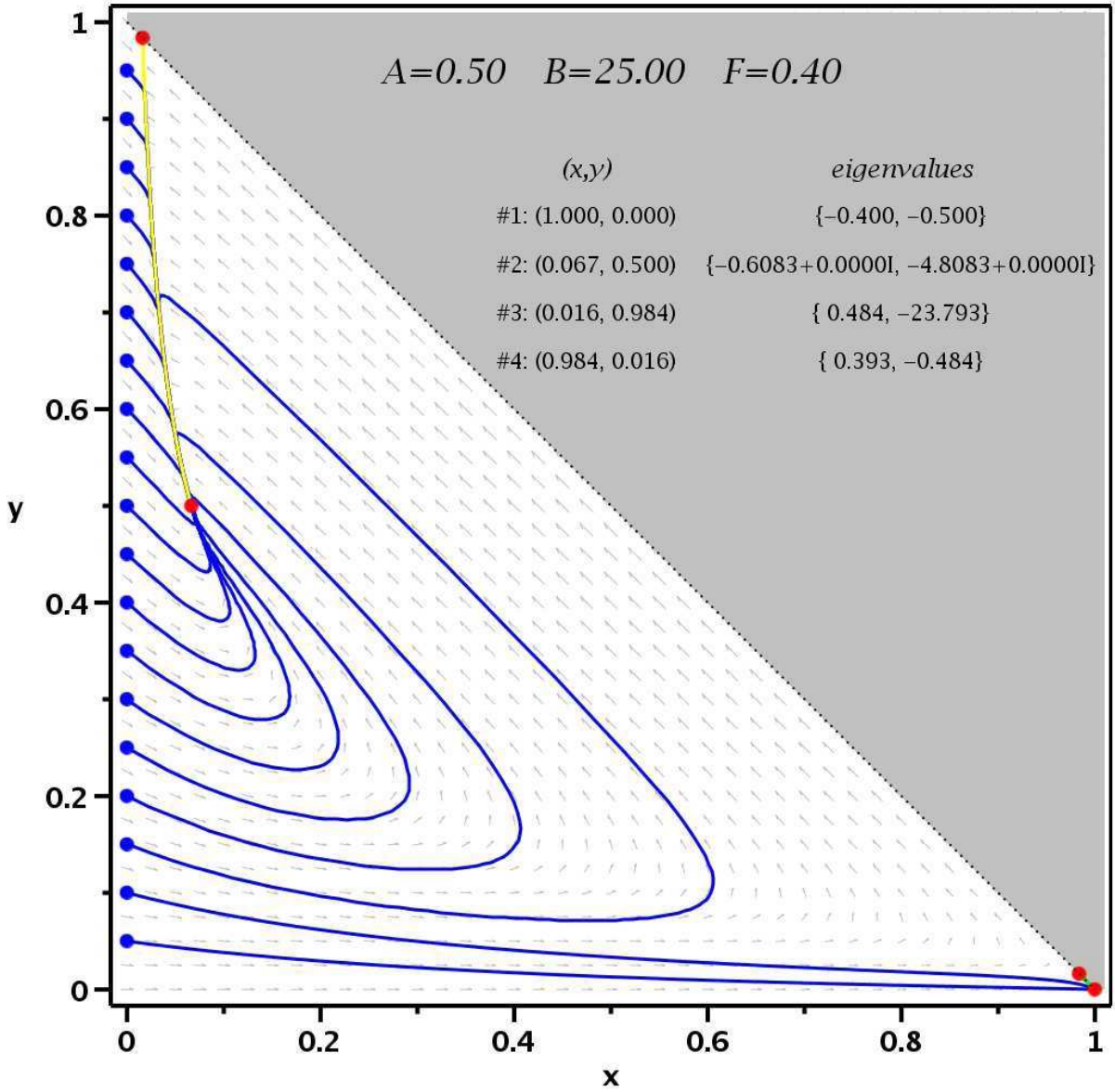}
\caption{Phase space plot for $A=0.5$, $B=25.$, $F=0.4$, as in Fig.~1,
  except that stationary point (ii) is here real, so the nearby phase space
  does not rotate.}
\label{Figure:3}
\end{figure}

\clearpage
\begin{figure}
\centering
\includegraphics[width=16cm]{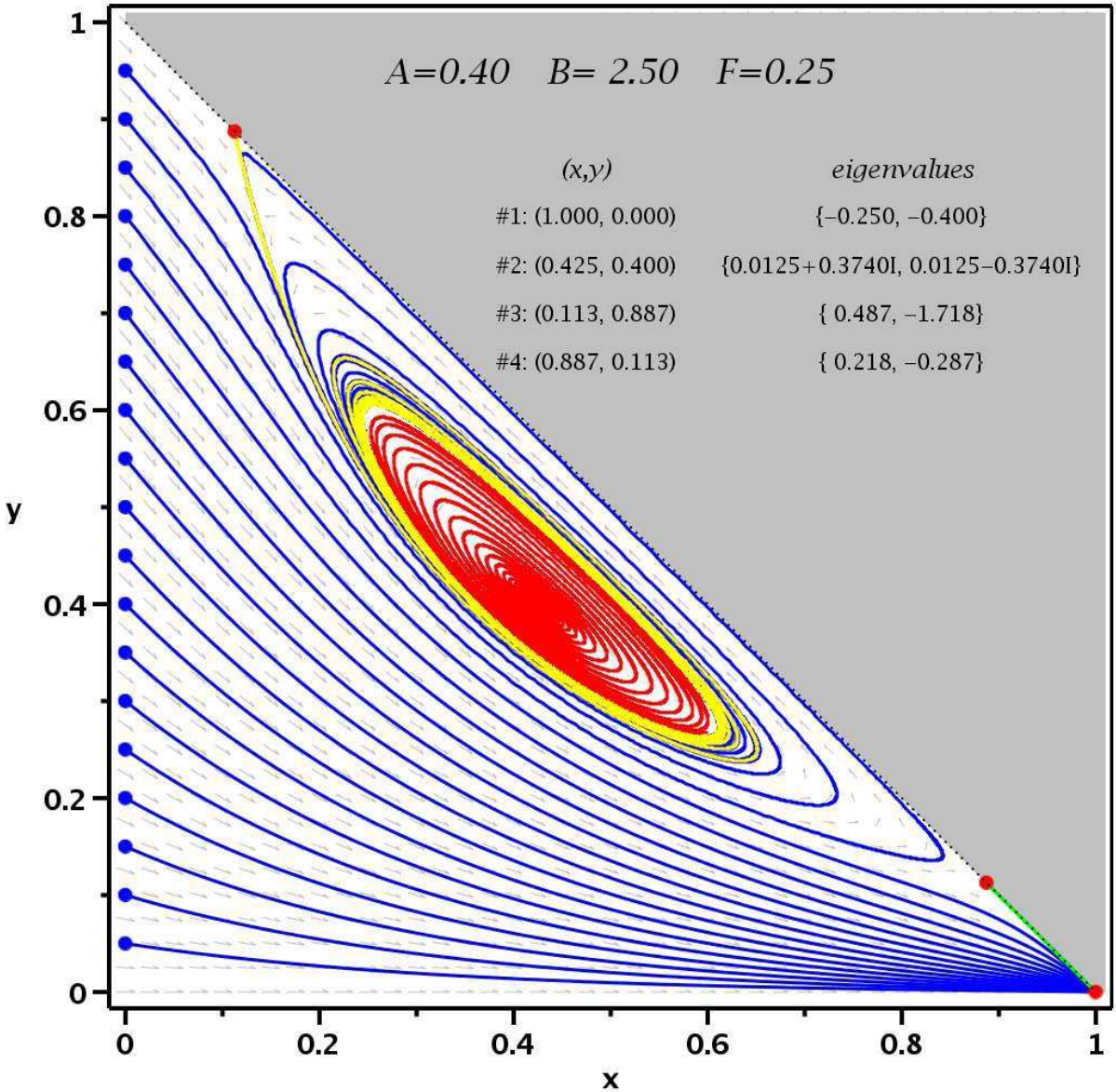}
\caption{Phase space plot for $A=0.4$, $B=2.5$, $F=0.25$, as in
  Fig.~1, except that stationary point (ii) is complex, but
  unstable. The red curve is a particular solution starting near point
  (ii) and spiraling outward, while the yellow curve spiral inward.  Both
  curves converge toward a stable limit cycle.  }
\label{Figure:4}
\end{figure}

\clearpage
\begin{figure}
\centering
\includegraphics*{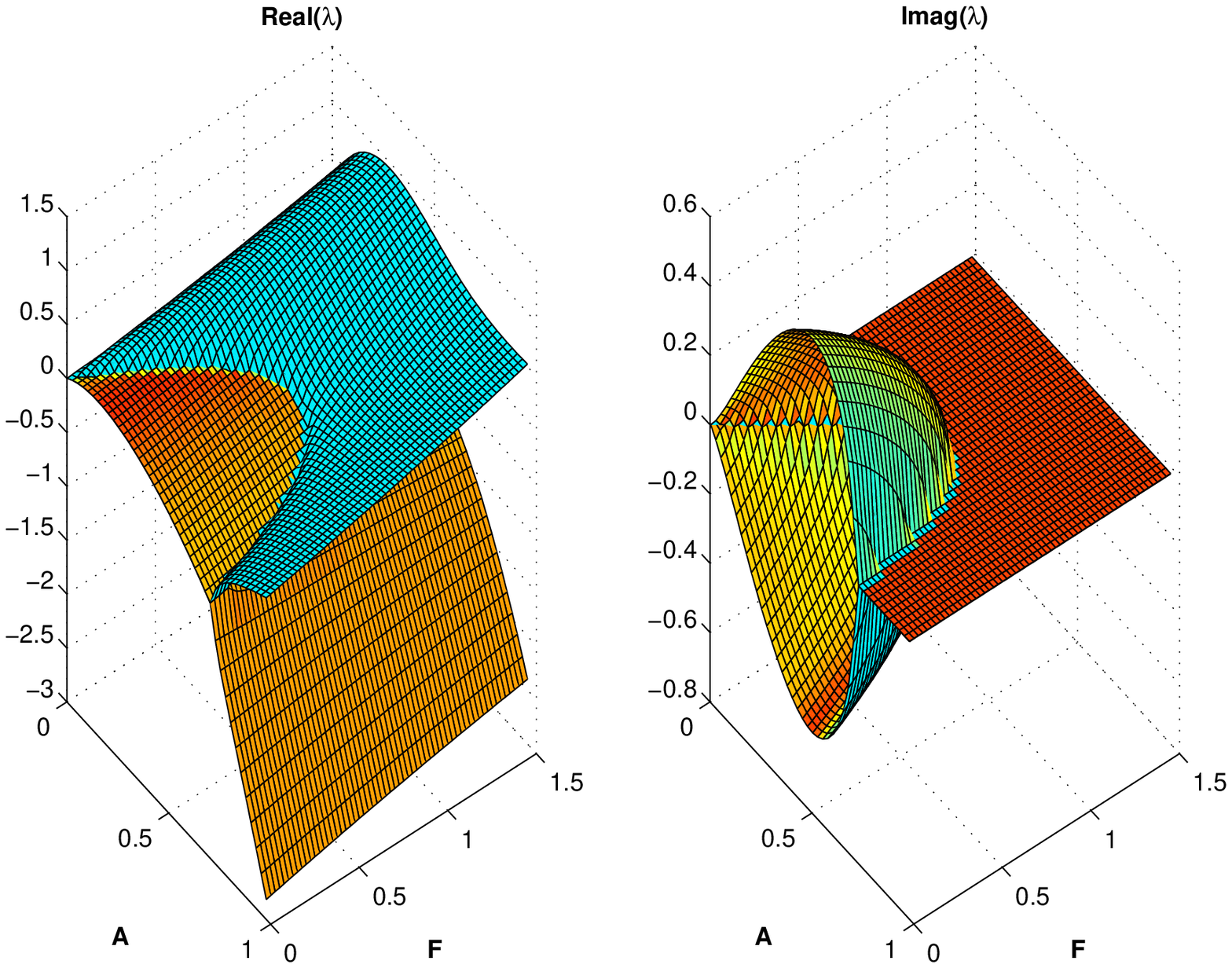}
\caption{Real (left) and imaginary (right) parts of eigenvalues
for
  the stationary point (ii) for $0 \le A \le1.0$, $B=3.0$, $0 \le F
  \le 1.5$.} \label{Figure:5}
\end{figure}

\clearpage
\begin{figure}
\centering
\includegraphics*{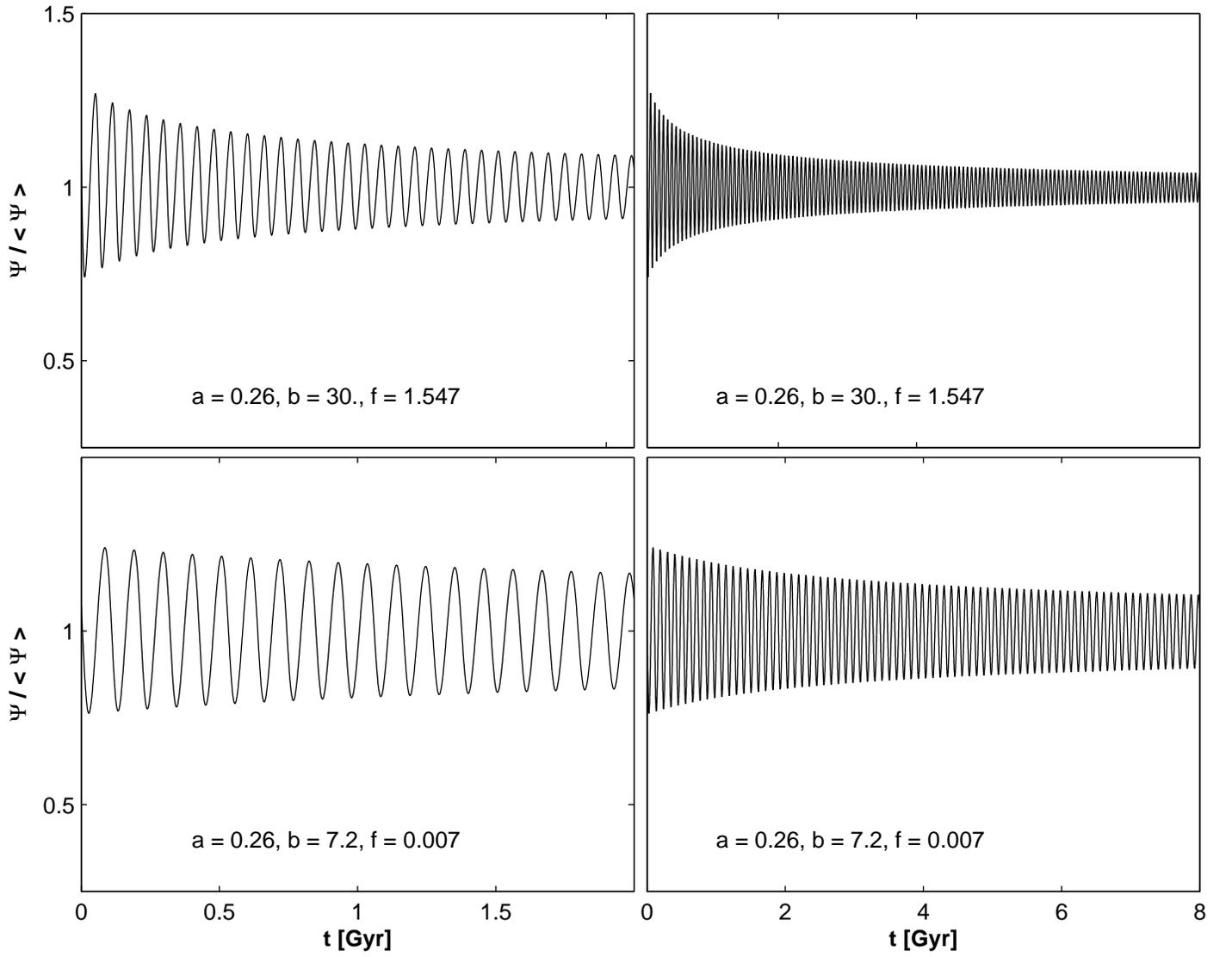}
\caption{Star formation rate (SFR) for giant (top) and dwarf
(bottom)
  galaxies.} \label{Figure:6}
\end{figure}

\clearpage
\begin{figure}
\centering
\includegraphics*{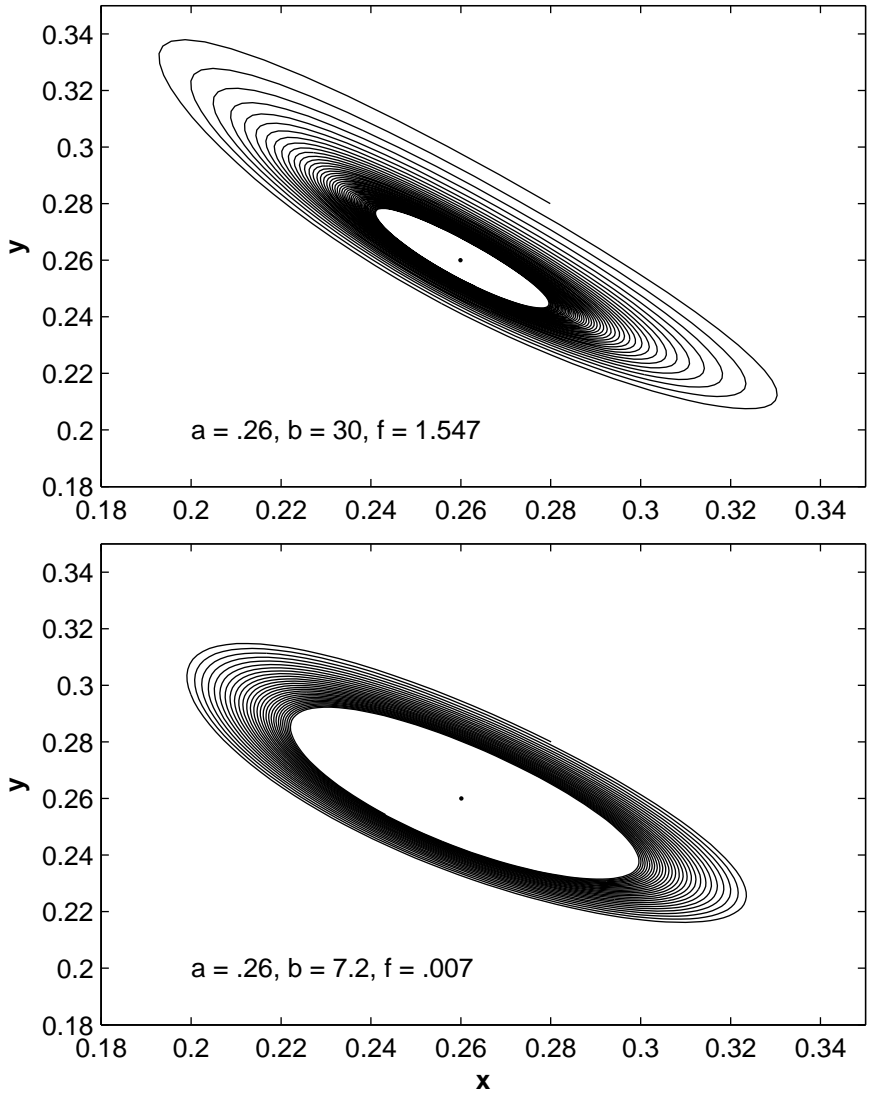}
\caption{Fast inspiraling behaviour for the stationary point (ii)
for higher value of $B$ ($\sim 30$) and slow behaviour for lower
value of  $B$ ($> 2.1$)} \label{Figure:7}
\end{figure}

\clearpage
\begin{figure}
\centering
\includegraphics*{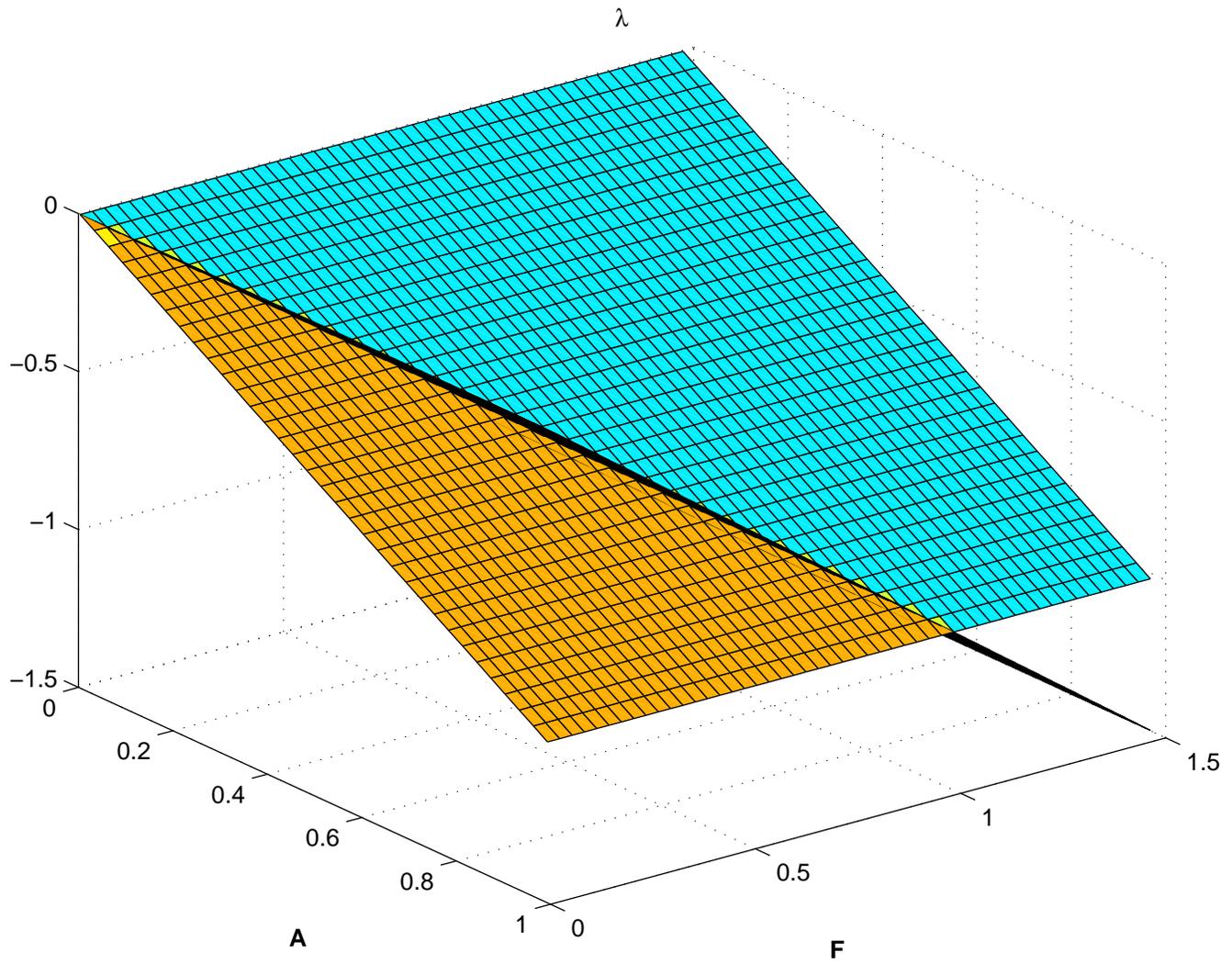}
\caption{Real parts of eigenvalues for the stationary point (i)
for $0 \le A \le1.0$, $B=3.0$, $0 \le F \le 1.5$.}
\label{Figure:8}
\end{figure}

\clearpage

\begin{figure}
\centering
\includegraphics*{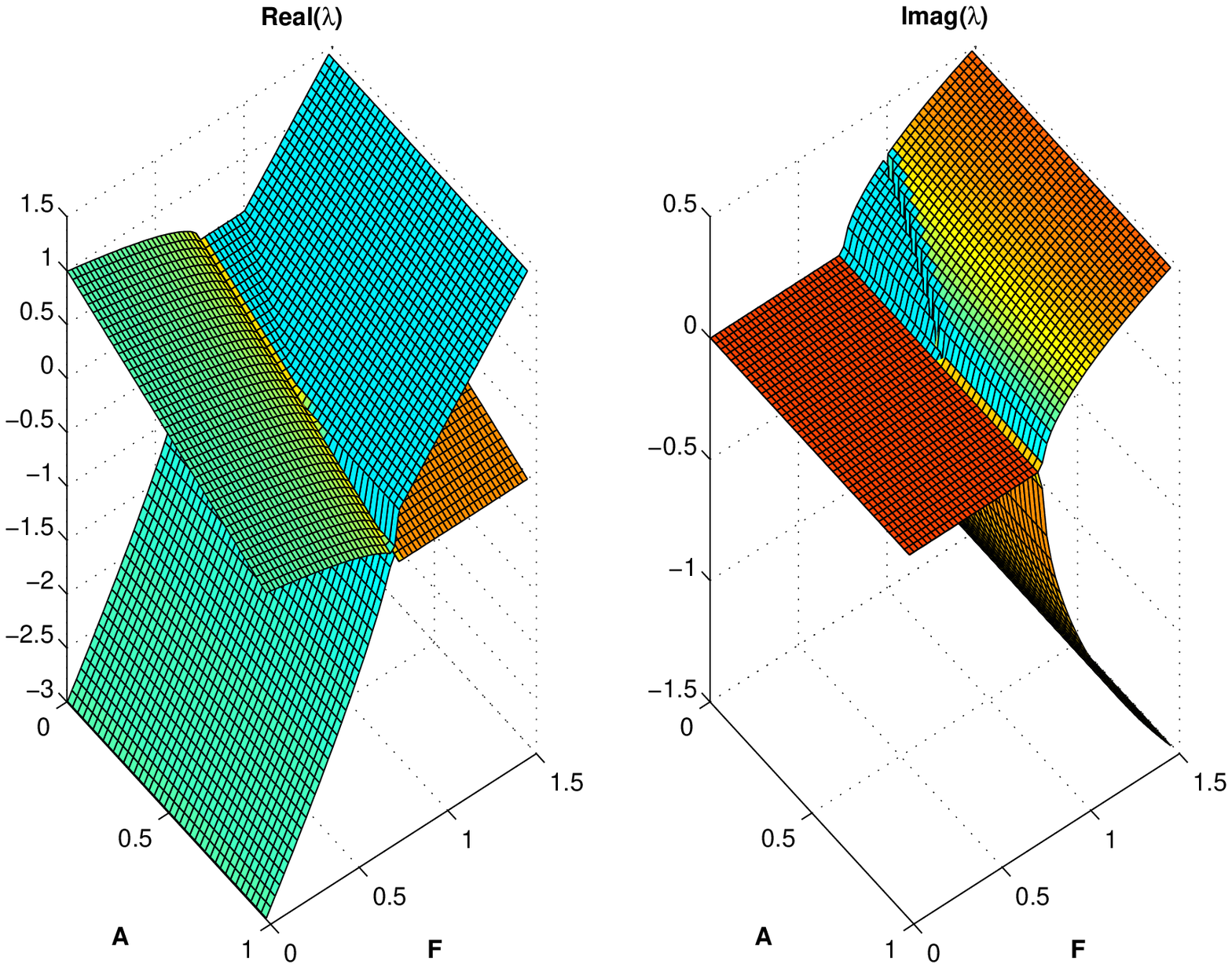}
\caption{Real (left) and imaginary (right) parts of eigenvalues
for the stationary point (iii) for $0 \le A \le1.0$, $B=3.0$, $0
\le F
  \le 1.5$.} \label{Figure:9}
\end{figure}

\clearpage

\begin{figure}
\centering
\includegraphics*{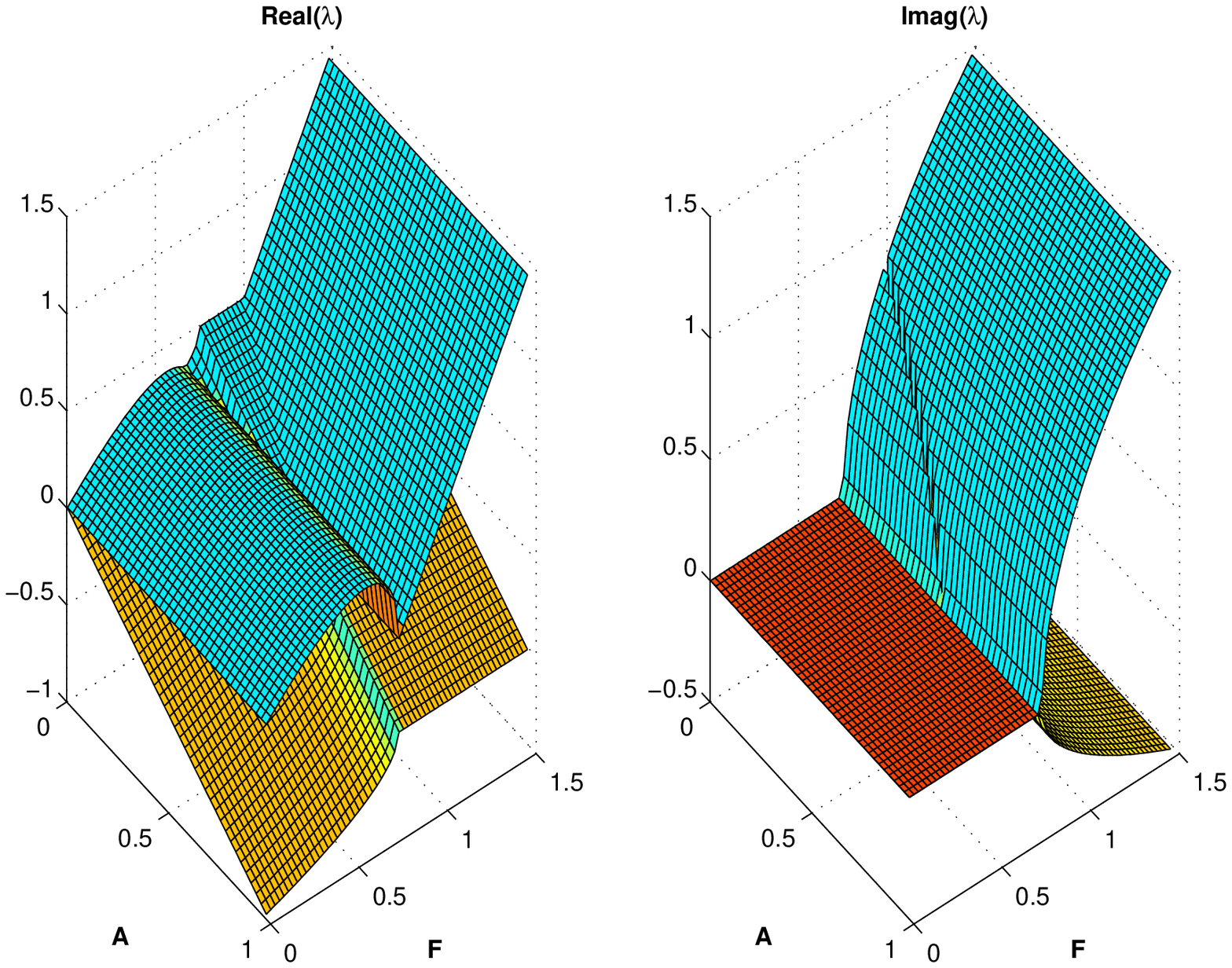}
\caption{Real (left) and imaginary (right) parts of eigenvalues
for the stationary point (iv) for $0 \le A \le1.0$, $B=3.0$, $0
\le F
  \le 1.5$.} \label{Figure:10}
\end{figure}

\clearpage

\begin{figure}
\centering
\includegraphics*{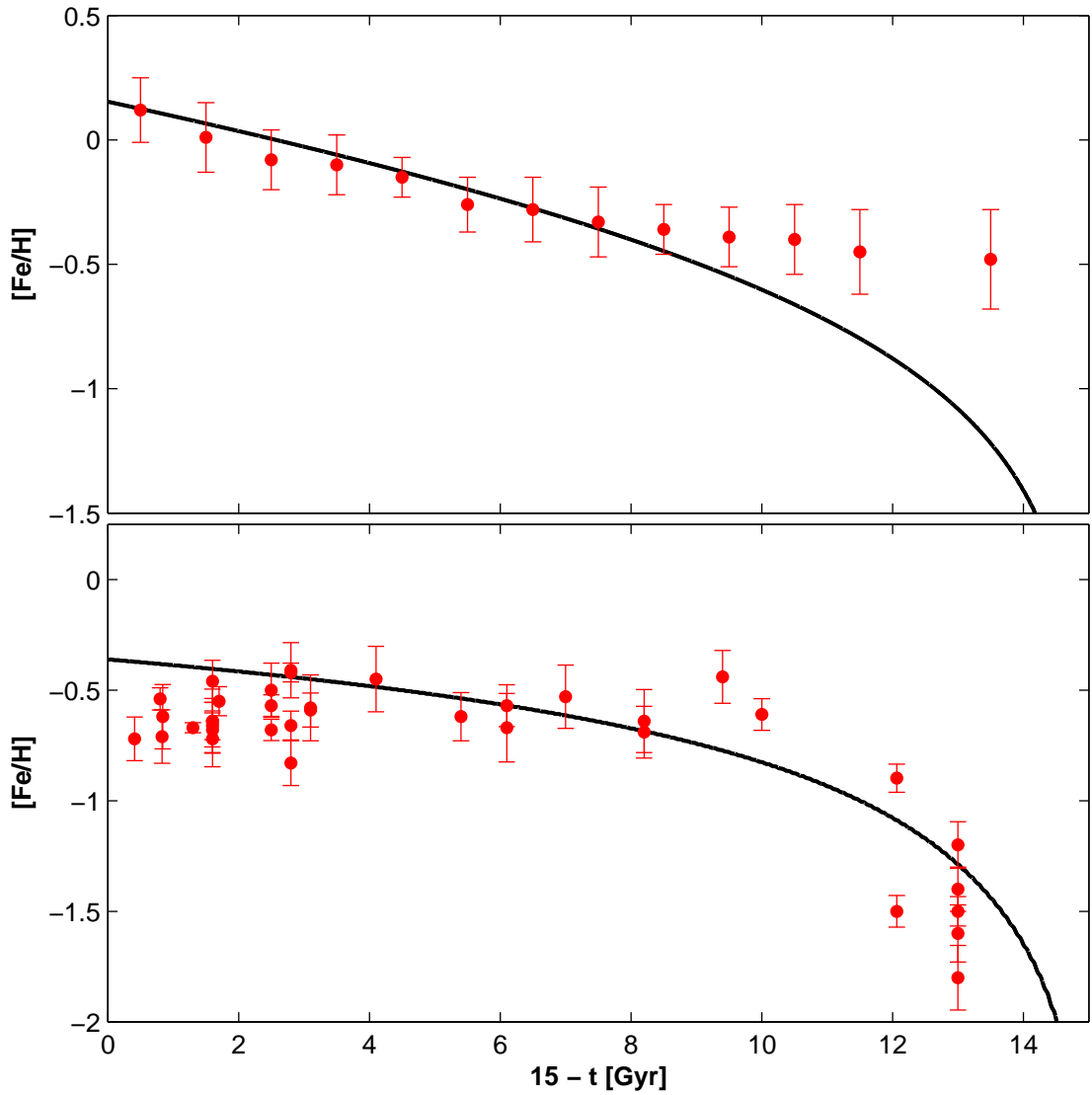}
\caption{Age-metallicity ([Fe/H]) diagram for giant (top) and
dwarf
  (bottom) galaxies for particular values of the input parameters. The
  red dots are observations from Rocha-Pinto et al.\ (2000) (top) and
  Chattopadhyay et al.\ (2012) (bottom).}
\label{Figure:11}
\end{figure}

\clearpage

\begin{figure}
\centering
\includegraphics*{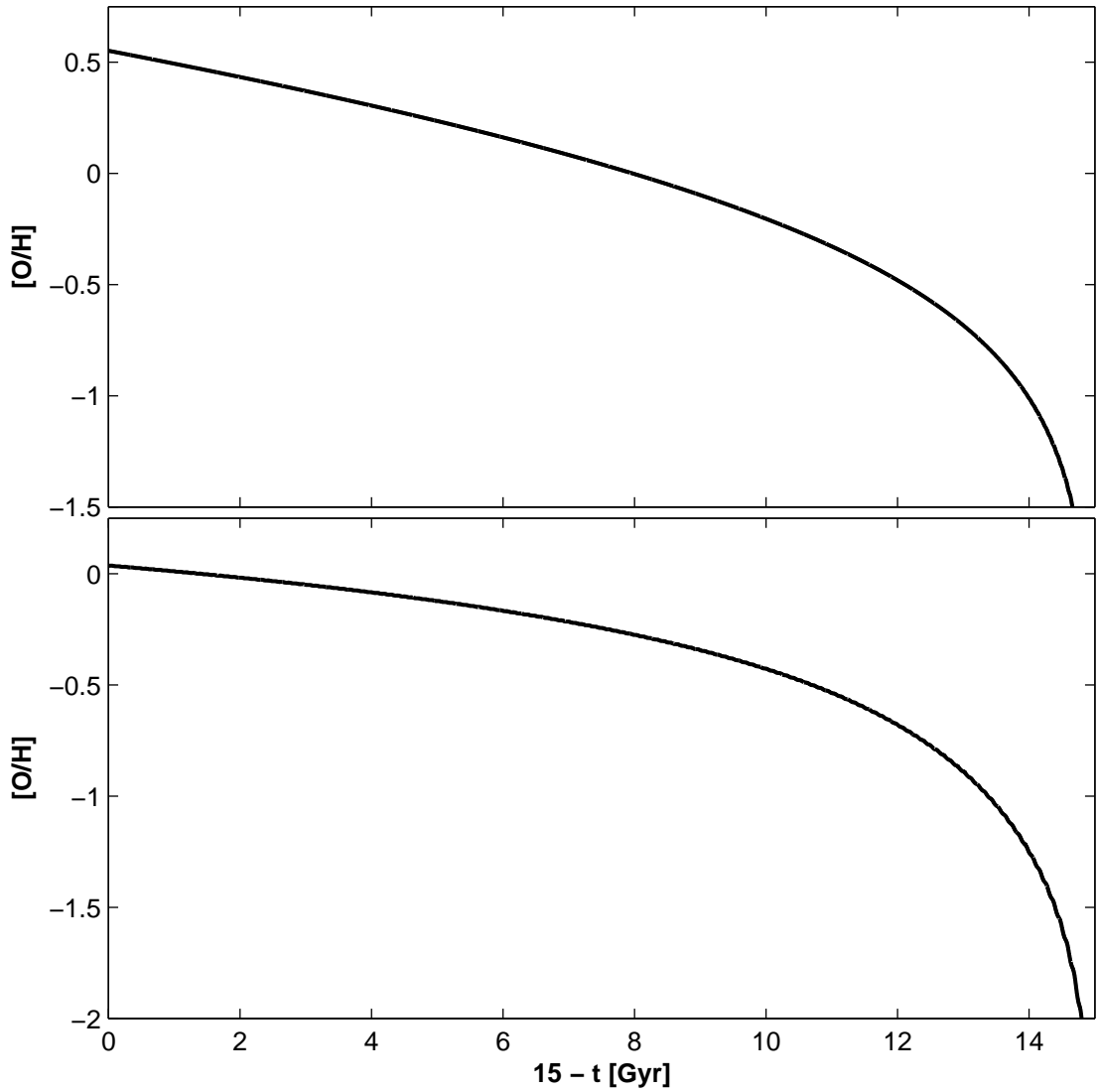}
\caption{Age-metallicity ([O/H]) diagram for giant (top) and dwarf
(bottom) galaxies for particular values of the input parameters. }
\label{Figure:12}
\end{figure}

\clearpage

\begin{table}
  \centering
  \caption{Mean SFR and abundances (at 15\,Gyr) in dwarf galaxies for
    various values of input parameters with $t_*=2.0\,$Gyr,
    $t_\mathrm{in} = 12.0\,$Gyr, $\frac{X_0^\mathrm{f}}{X_{0,\odot}} =
    \frac{X_\mathrm{Fe}^\mathrm{f}}{X_{\mathrm{Fe},\odot}} = 0.0$,
    $R_\mathrm{ins} = 0.16$, $Y_{0,\mathrm{ins}}/X_{0,\odot}= 0.70$,
    $Y_{\mathrm{Fe},\mathrm{ins}}/X_{\mathrm{Fe},\odot} = 0.28$,
    $v_\mathrm{c} = 20 \rm km\, s^{-1}$, $v_{c0} = 100 \rm
    km\,s^{-1}$, $\alpha = 1.2$ }
\begin{tabular}{c c c c c c c}
\hline\hline\\
$A$ & $B$ & $F$ & $\langle \frac{\psi t_*}{M_\mathrm{g}}\rangle$ &
$\langle \frac{\psi t_*}{M_\mathrm{g}}\rangle_\sigma$ & $\mathrm{[O/H]}_\mathrm{max}$ & $\mathrm{[Fe/H]}_\mathrm{max}$ \\
\hline
0.260& 7.50 & 0.016 & 0.2564 & 0.00013028 & 0.0350 & -0.3630 \\
0.260& 7.51 & 0.019 & 0.2572 & 0.00015528 & 0.0354 & -0.3625 \\
0.260& 7.52 & 0.022 & 0.2581 & 0.00019248 & 0.0358 & -0.3621 \\
\hline
0.265& 7.50 & 0.016 & 0.2514 & 0.00004227 & 0.0323 & -0.3656 \\
0.265& 7.51 & 0.019 & 0.2522 & 0.00004468 & 0.0328 & -0.3652 \\
0.265& 7.52 & 0.022 & 0.2529 & 0.00004736 & 0.0332 & -0.3647 \\
\hline
0.270& 7.50 & 0.016 & 0.2466 & 0.00002454 & 0.0295 & -0.3684 \\
0.270& 7.51 & 0.019 & 0.2474 & 0.00002538 & 0.0300 & -0.3679 \\
0.270& 7.52 & 0.022 & 0.2481 & 0.00002627 & 0.0305 & -0.3675 \\
\hline
0.305& 7.50 & 0.016 & 0.2162 & 0.00000516 & 0.0077 & -0.3902 \\
0.305& 7.51 & 0.019 & 0.2169 & 0.00000521 & 0.0083 & -0.3896 \\
0.305& 7.52 & 0.022 & 0.2176 & 0.00000526 & 0.0089 & -0.3890 \\
\hline \hline
\end{tabular}
\end{table}

\begin{table}
  \centering
  \caption{Values of the parameters and initial values of the
    abundances.}
\begin{tabular}{@{}lc}
\hline\hline
Parameter & values   \\
\hline
$A$ & $0 - 1.0$ \\
$B$ & $0 - 100$ \\
$Y_{0,\mathrm{ins}}$ (in $X_{0\odot}$) & 0.70 \\
$Y_{\mathrm{Fe},\mathrm{ins}}$ (in $X_{\mathrm{Fe}\odot}$) & 0.28 \\
$X_0^\mathrm{f}$ (in $X_{\mathrm{Fe}\odot})$ & 0.0 \\
$X_{Fe}^\mathrm{f}$ (in $X_{\mathrm{Fe}\odot})$ & 0.0\\
$R_\mathrm{ins}$ & 0.16 \\
$t_*$ (in Gyr) & 1.0, 2.0 \\
$t_\mathrm{in}$ (in Gyr) & 2.0, 12.0 \\
$v_\mathrm{c}$ (in $\rm km\,s^{-1}$) & 200, 20 \\
$v_{c0}$ (in $\rm km\,s^{-1}$) & 100 \\
$\alpha$ & 1.2 \\
\hline\hline
\end{tabular}
\end{table}

\clearpage

\begin{table}
\centering \caption{Mean SFR and abundances (at 15\,Gyr) in giant
  galaxies for various values of input parameters with $t_* = 1.0\,$Gyr,
  $t_\mathrm{in} = 2.0\,$Gyr,
  $\frac{X_0^\mathrm{f}}{X_{0,\odot}} = \frac{X_\mathrm{Fe}^\mathrm{f}}{X_{\mathrm{Fe},\odot}} = 0.0$,
  $R_\mathrm{ins} = 0.16$,
  $Y_{0,\mathrm{ins}}/X_{0,\odot} = 0.70$, $Y_{\mathrm{Fe},\mathrm{ins}}/X_{\mathrm{Fe},\odot} = 0.28$,
  $v_\mathrm{c} = 200\, \rm km\,s^{-1}$, $v_{c0} = 100\, \rm km\,s^{-1}$, $\alpha = 1.2$ }
\begin{tabular}{c c c c c c c}
\hline\hline\\
$A$ & $B$ & $F$ & $\langle \frac{\psi t_*}{M_\mathrm{g}}\rangle $
&
$\langle \frac{\psi t_*}{M_\mathrm{g}}\rangle_\sigma$ & $\mathrm{[O/H]}_\mathrm{max}$ & $\mathrm{[Fe/H]}_\mathrm{max}$ \\
\hline
0.260& 30.1& 1.55&    0.25969&   0.00010330 &     0.55039 &   0.15245 \\
0.260& 30.2& 1.56&    0.26018&   0.00014741 &     0.55128 &   0.15334 \\
0.260& 30.3& 1.57&    0.26085&   0.00025096 &     0.55262 &   0.15468 \\
\hline
0.265& 30.1& 1.64&    0.26483&   0.00011325 &     0.56037 &   0.16243 \\
0.265& 30.2& 1.65&    0.26531&   0.00016407 &     0.56123 &   0.16329 \\
0.265& 30.3& 1.66&    0.26594&   0.00027099 &     0.56245 &   0.16451 \\
\hline
0.270& 30.1& 1.73&    0.26977&   0.00010806 &     0.56969 &   0.17175 \\
0.270& 30.2& 1.74&    0.27017&   0.00014613 &     0.57038 &   0.17244 \\
0.270& 30.3& 1.75&    0.27069&   0.00022566 &     0.57133 &   0.17339 \\
\hline
0.275& 30.1& 1.82&    0.27451&   0.00009238 &     0.57843 &   0.18049 \\
0.275& 30.2& 1.83&    0.27483&   0.00011220 &     0.57895 &   0.18101 \\
0.275& 30.3& 1.84&    0.27519&   0.00014867 &     0.57955 &   0.18161 \\
\hline\hline
\end{tabular}
\end{table}

\clearpage

\begin{table}
  \centering
  \caption{Birth rates in the last 100, 500, 1000\,Myr and average
    duty cycles for giant galaxies for various values of input
    parameters.}
\begin{tabular}{c c c c c c c c}
\hline\hline\\
$A$ & $B$ & $F$ &$\langle\frac{\psi t_*}{M_\mathrm{g}}\rangle$&$b_{100}$ & $b_{500}$ & $b_{1000}$ & average cycle \\
    &     &     &                                           &         &          &           &(in $10^7$yr)  \\
\hline
0.260& 30.1&   1.55&   0.26041 &   0.99957&   0.99910&   0.99918&   0.84750 \\
0.260& 30.2&   1.56&   0.26099 &   1.00020&   0.99844&   0.99898&   1.99867 \\
0.260& 30.3&   1.57&   0.26160 &   1.00791&   1.00043&   0.99962&   2.98560 \\
\hline
0.265& 30.1&   1.64&   0.26564 &   0.99979&   0.99930&   0.99907&   1.13534 \\
0.265& 30.2&   1.65&   0.26620 &   1.00239&   0.99973&   0.99889&   2.21000 \\
0.265& 30.3&   1.66&   0.26680 &   0.99424&   0.99977&   0.99844&   3.27832 \\
\hline
0.270& 30.1&   1.73&   0.27046 &   0.99875&   0.99915&   0.99913&   0.92132 \\
0.270& 30.2&   1.74&   0.27094 &   0.99894&   0.99933&   0.99885&   2.10058 \\
0.270& 30.3&   1.75&   0.27145 &   0.99408&   0.99883&   0.99880&   2.94427 \\
\hline
0.275& 30.1&   1.82&   0.27516 &   0.99916&   0.99926&   0.99932&   0.45329 \\
0.275& 30.2&   1.83&   0.27557 &   0.99847&   0.99903&   0.99916&   1.11802 \\
0.275& 30.3&   1.84&   0.27599 &   0.99812&   0.99905&   0.99904&   2.04222 \\
\hline
0.280& 30.1&   1.91&   0.27964 &   0.99952&   0.99950&   0.99950&   0.23935 \\
0.280& 30.2&   1.92&   0.27996 &   0.99937&   0.99939&   0.99940&   0.35090 \\
0.280& 30.3&   1.93&   0.28030 &   0.99903&   0.99923&   0.99930&   0.55776 \\
\hline
0.305& 30.1&   2.36&   0.30009 &   0.99999&   0.99999&   0.99999&   0.11861 \\
0.305& 30.2&   2.37&   0.30018 &   0.99999&   0.99999&   0.99999&   0.12270 \\
0.305& 30.3&   2.38&   0.30026 &   0.99999&   0.99999&   0.99999&   0.11878 \\
\hline \hline
\end{tabular}
\end{table}

\begin{table}
  \centering
  \caption{Birth rates in the last 100, 500, 1000 Myr and average duty
    cycles for dwarf galaxies for various values of input parameters
    and compared with the observed birth rates of dwarf irregular
    galaxies (Weisz et al.\ 2008).}
\begin{tabular}{c c c c c c c c c c}
  \hline\hline\
  $A$ & $B$ & $F$ & $b_{100}$   & $b_{100}$              & $b_{500}$  & $b_{500}$ & $b_{1000}$  & $b_{1000}$& average cycles.\\
   &     &     &(predicted) &(observed, Galaxy name)&(predicted)&(observed)&(predicted)&(observed)& (in $10^7$ yr) \\
\hline
0.260& 7.50 &  0.016 & 0.99951 & 0.37 (Garland)  & 0.99951 & 0.25 & 0.99950 & 0.15 & 0.44958 \\
0.260& 7.51 &  0.019 & 0.99939 & 1.84 (Dwarf A)  & 0.99941 & 1.08 & 0.99939 & 0.92 & 0.68348 \\
0.260& 7.52 &  0.022 & 0.99924 & 1.24 (DDO 53)   & 0.99932 & 0.76 & 0.99931 & 1.08 & 1.48730 \\
\hline
0.265& 7.50 &  0.016 & 1.00002 & 1.38 (IC 2574)  & 1.00002 & 1.38 & 1.00002 & 1.97 & 0.14765 \\
0.265& 7.51 &  0.019 & 1.00000 & 1.73 ( WLM )    & 1.00000 & 2.08 & 1.00000 & 2.24 & 0.13765 \\
0.265& 7.52 &  0.022 & 0.99999 & 1.61 (IC 1613)  & 0.99999 & 0.95 & 0.99999 & 0.83 & 0.13234 \\
\hline
0.270& 7.50 &  0.016 & 1.00012 & 1.36 (NGC 3109) & 1.00012 & 1.05 & 1.00012 & 0.75 & 0.11794 \\
0.270& 7.51 &  0.019 & 1.00012 & 0.34 (IC 10)    & 1.00012 & 1.05 & 1.00012 & 0.81 & 0.11612 \\
0.270& 7.52 &  0.022 & 1.00011 & --              & 1.00011 &  --  & 1.00011 & --   & 0.11638 \\
\hline\hline
\end{tabular}
\end{table}

\clearpage

\begin{table}
\centering \caption{Some parameter values where a limit cycle is found.}

\begin{tabular}{@{}rrr}
\hline \hline
$A$ & $B$ & $F$ \\
\hline
0.10 &  63.00 &  0.00 \\
0.10 & 100.00 &  0.25 \\
0.10 & 100.00 &  0.40 \\
\hline
0.15 &  16.00 &  0.00 \\
0.15 &  25.00 &  0.00 \\
0.15 &  25.00 &  0.10 \\
0.15 &  40.00 &  0.25 \\
0.15 &  40.00 &  0.40 \\
0.15 & 100.00 &  1.60 \\
\hline
0.20 &   6.30 &  0.00 \\
0.20 &  10.00 &  0.00 \\
0.20 &  10.00 &  0.10 \\
0.20 &  16.00 &  0.10 \\
0.20 &  16.00 &  0.16 \\
0.20 &  16.00 &  0.25 \\
\hline
0.25 &   2.50 &  0.00 \\
0.25 &   4.00 &  0.00 \\
0.25 &   6.30 &  0.00 \\
0.25 &   6.30 &  0.10 \\
0.25 &   6.30 &  0.16 \\
0.25 &  10.00 &  0.16 \\
0.25 &  10.00 &  0.25 \\
0.25 &  16.00 &  0.63 \\
\hline
0.30 &   1.00 &  0.00 \\
0.30 &   1.60 &  0.00 \\
0.30 &   2.50 &  0.00 \\
0.30 &   2.50 &  0.10 \\
0.30 &   4.00 &  0.00 \\
0.30 &   4.00 &  0.10 \\
0.30 &   4.00 &  0.16 \\
\hline \hline
\end{tabular}
\begin{tabular}{@{}rrr}
\hline \hline
$A$ & $B$ & $F$ \\
\hline
0.30 &   5.00 &  0.10 \\
0.30 &   6.30 &  0.25 \\
0.30 &  10.00 &  0.63 \\
\hline
0.35 &   0.16 &  0.00 \\
0.35 &   0.25 &  0.00 \\
0.35 &   0.40 &  0.00 \\
0.35 &   0.63 &  0.00 \\
0.35 &   1.00 &  0.00 \\
0.35 &   1.60 &  0.00 \\
0.35 &   1.60 &  0.10 \\
0.35 &   2.50 &  0.10 \\
0.35 &   2.50 &  0.16 \\
0.35 &   4.00 &  0.25 \\
0.35 &  10.00 &  1.00 \\
\hline
0.40 &   0.10 &  0.00 \\
0.40 &   0.16 &  0.00 \\
0.40 &   0.25 &  0.00 \\
0.40 &   0.40 &  0.00 \\
0.40 &   0.63 &  0.00 \\
0.40 &   1.00 &  0.00 \\
0.40 &   1.00 &  0.10 \\
0.40 &   1.60 &  0.10 \\
0.40 &   1.60 &  0.16 \\
0.40 &   2.50 &  0.25 \\
\hline
0.45 &   0.10 &  0.00 \\
0.45 &   0.16 &  0.00 \\
0.45 &   0.25 &  0.00 \\
0.45 &   0.40 &  0.00 \\
0.45 &   0.63 &  0.10 \\
0.45 &   1.00 &  0.16 \\
\vspace{-1mm}\\
\hline \hline
\end{tabular}
\end{table}

\end{document}